\pdfoutput=1 

\documentclass[iop]{emulateapj}

\usepackage{graphicx}
\usepackage{epstopdf}

\newcommand{\kms}{\hbox{km s$^{-1}$}}
\newcommand{\msun}{\hbox{$M_\odot$}}
\newcommand{\re}{\hbox{$R_{\rm e}$}}
\newcommand{\remaj}{\hbox{$R_{\rm e}^{\rm maj}$}}
\newcommand{\reffig}[1]{Fig.~\ref{#1}}
\newcommand{\atl}{ATLAS$^{\rm 3D}$}

\slugcomment{The Astrophysical Journal Letters. Received 2013 September 4; accepted 2013 September 27}
\shorttitle{Galaxies mass-size in densest environment}
\shortauthors{Cappellari}

\begin{document}

\title{Effect of environment on galaxies mass-size distribution:\\ unveiling the transition from outside-in to inside-out evolution}

\author{Michele Cappellari}

\affil{Sub-department of Astrophysics, Department of Physics, University of Oxford, Denys Wilkinson Building, Keble Road, Oxford, OX1~3RH, UK}

\begin{abstract}
The distribution of galaxies on the mass-size plane as a function of redshift or environment is a powerful test for galaxy formation models. Here we use integral-field stellar kinematics to interpret the variation of the mass-size distribution in two galaxy samples spanning extreme environmental densities. The samples are both identically and nearly mass-selected (stellar mass $M_\ast\ga6\times10^9$ \msun) and volume-limited. The first consists of nearby field galaxies from the \atl\ parent sample. The second consists of galaxies in the Coma Cluster (Abell 1656), one of densest environments for which good resolved spectroscopy can be obtained. The mass-size distribution in the dense environment differs from the field one in two ways: (i) spiral galaxies are replaced by bulge-dominated disk-like fast-rotator early-type galaxies (ETGs), which follow the {\em same} mass-size relation and have the {\em same} mass distribution as in the field sample; (ii) the slow rotator ETGs are segregated in mass from the fast rotators, with their size increasing proportionally to their mass. A transition between the two processes appears around the stellar mass $M_{\rm crit}\approx2\times10^{11}$ \msun. We interpret this as evidence for bulge growth (outside-in evolution) and bulge-related environmental quenching dominating at low masses, with little influence from merging, while significant dry mergers (inside-out evolution) and halo-related quenching driving the mass and size growth at the high-mass end. The existence of these two processes naturally explains the diverse size evolution of galaxies of different masses and the separability of mass and environmental quenching.
\end{abstract}


\section{Introduction}

Galaxy stellar masses and sizes are powerful observables to study galaxy evolution. They vary with time or environment, during the hierarchical galaxy growth. But growth rates depend on the assembly process \citep[e.g.][]{Khochfar2006,Naab2009,Hopkins2010}. A theme emerging from both theory and observations is a dichotomy between the redshift and environmental evolution of galaxy sizes as a function of stellar mass.

On one hand the passive early-type galaxies (ETGs) with stellar masses $M_\ast\ga10^{11}$ \msun\ are found to be, on {\em average}, smaller and denser at redshift $z\sim2$. This results comes from both photometry \citep[e.g.][]{Daddi2005,Trujillo2006gems,vanDokkum2008} and stellar velocity dispersion \citep[e.g.][]{Cappellari2009,Cenarro2009,vandeSande2013}. On the other hand lower-mass disks, show no significant size evolution out to $z\sim1$ \citep[e.g.][]{Barden2005,Sargent2007} and little evolution out to $z\sim3$ \citep{Nagy2011}.

A dichotomy is also seen in the evolution of galaxy profiles. Massive passive galaxies (present day $M_\ast\approx3\times10^{11}$ \msun) build their mass mostly inside-out, by gradually assembling a stellar halo around a compact spheroid \citep{vanDokkum2010profiles}. In contrast, lower mass star forming systems (present day $M_\ast\approx5\times10^{10}$ \msun) indicate and early bulge growth followed by a modest but uniform growth in size at all radii \citep{vanDokkum2013}.

An increase in the galaxies number density has a similar effect on galaxy properties as time evolution. This is likely because at high redshift the abundance of massive halos declines and fewer galaxies are in clusters than locally. Overdensities transform spirals into passive ETGs \citep{Dressler1980} and increase mean galaxy masses \citep{Kauffmann2004}.
In apparent contrast to redshift evolution however, both spirals and ETGs follow nearly the same mass-size relations in different environments \citep[e.g.][]{Maltby2010,Huertas-Company2013,Poggianti2013}. Although some size differences were reported at $z\sim1$, they are either small (25\% in \citealt{Cooper2012}), or only affect massive galaxies ($M_\ast\ga2\times10^{11}$ \msun\ in \citealt{Lani2013}).

Here we want to understand the origin of the observed trends, using the exquisite detail on the fossil record of galaxy formation one can obtain only for nearby galaxies. We combine available integral-field stellar kinematics and Hubble Space Telescope (HST) photometry of the inner surface brightness profiles to robustly recognize the merger history the galaxies have experienced. We study two extreme environments which differ by almost three orders of magnitude in galaxy number density. We show that a simple picture can reconcile the trends of galaxy sizes with the detailed fossil record of galaxy formation.

\section{Sample and data}

\subsection{Selection}

The galaxies in both our low-density and high-density samples are identically selected from the 2MASS \citep{Skrutskie2006} Extended Source Catalog (XSC) for having total absolute magnitude  (from XSC keyowrd k\_m\_ext) $M_{K_s}<-21.5$ mag ($M_\ast\ga6\times10^9$ \msun). 

Our low-density (field) sample consists of all 743 galaxies in the \atl\ {\em parent} sample \citep{Cappellari2011a} not belonging to the Virgo cluster according to that paper. This provides a clean group/field environment as shown in \citet{Cappellari2011b}. The sample is volume-limited within a distance of 42 Mpc and 98\% complete \citep[sec.~2.1 of][]{Cappellari2013p20}. The median number density of this sample is $\log\Sigma_3=-0.5$ (Mpc$^{-2}$) using the 3rd nearest neighbor estimator determinations from \citet{Cappellari2011b}.

Our high-density (cluster) sample consists of all 160 cluster-member galaxies within a circle of area 1 deg$^2$ centered on the core of the Coma cluster (Abell 1656). The center was assumed as the midpoint between the two central galaxies NGC~4874 and NGC~4884 which nearly coincides with the peak of the X-ray emission \citep{White1993}. We adopted a Coma cluster distance of 100 Mpc \citep[see][]{Carter2008}. At our luminosity the XSC is essentially complete but, after visual inspection of SDSS images of the cluster, we manually added the five galaxies NGC~4871, NGC~4872, NGC~4882 ($=$NGC~4886), PGC~044644 and PGC~044651, which were likely missed for lying within the stellar halo of the central galaxies.

All the 2MASS selected galaxies in Coma have a redshift in either SDSS or the NASA Extragalactic Database. We removed the only 12 galaxies with recession velocity $V_{\rm hel}>11000$ \kms, providing a complete sample of cluster-member galaxies within the given cylinder. The selection radius of 1.0 Mpc is about 1/3 of the dark halo virial radius \citep{Lokas2003}. The Coma sample has a median density $\log\Sigma_3=2.0$ (Mpc$^{-2}$), when computed in identical manner as for \atl.

\begin{figure}
\centering
\includegraphics[height=6cm]{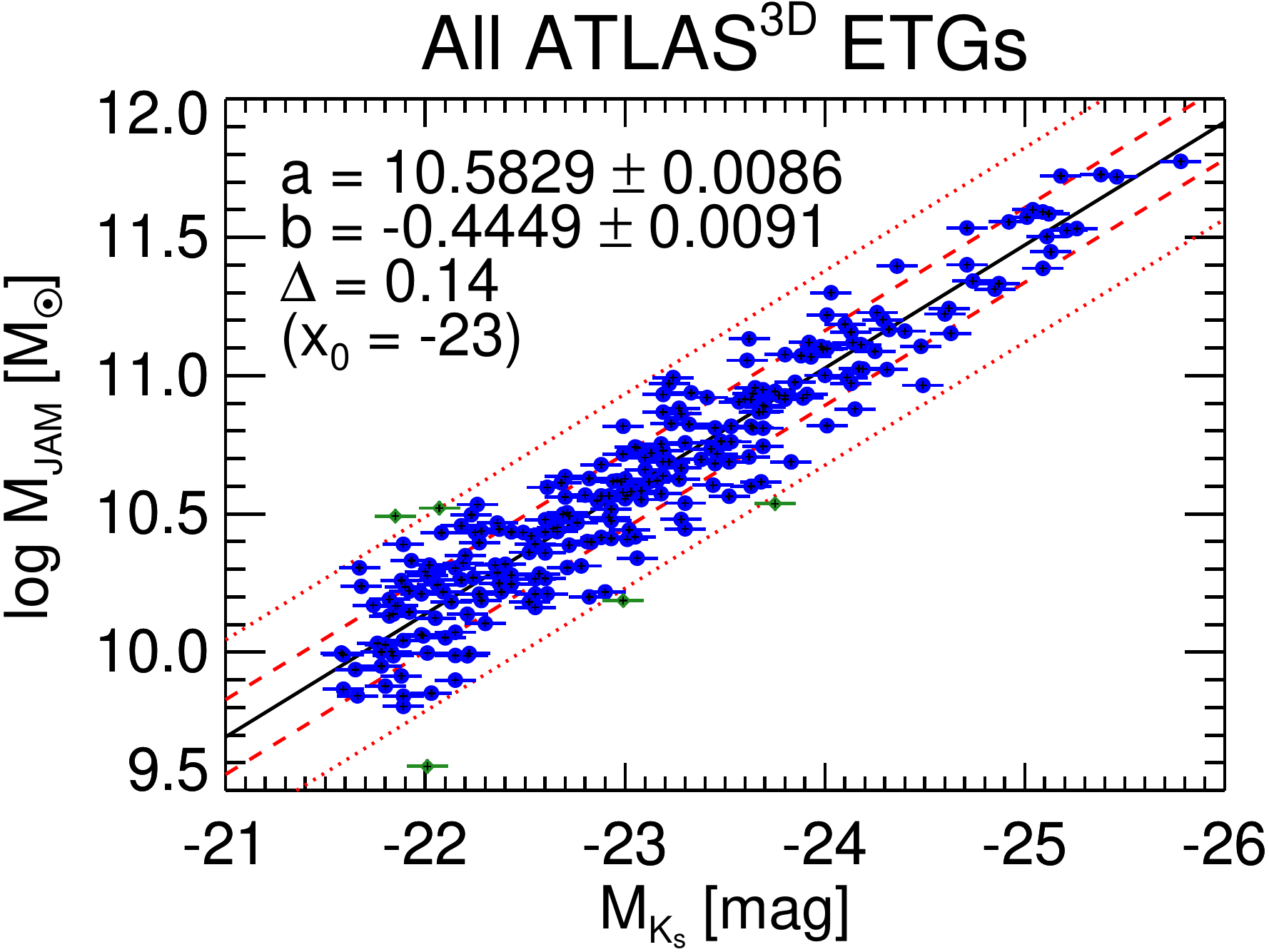}
\caption{From $K_s$ luminosity to stellar mass. The dynamical mass $M_{\rm JAM}\approx M_\ast$ is plotted against the total luminosity $M_{K_s}$. The best fitting relation (solid line) and the $1\sigma$ and $2.6\sigma$ bands are also shown with the dashed and dotted lines respectively. The green filled diamonds were automatically excluded from the fit. The best-fitting linear parameter and errors are printed, together with the observed scatter $\Delta$ and pivot magnitude $x_0$.}
\label{fig:from_kmag_to_mass}
\end{figure}

\begin{figure}
\centering
\includegraphics[height=6cm]{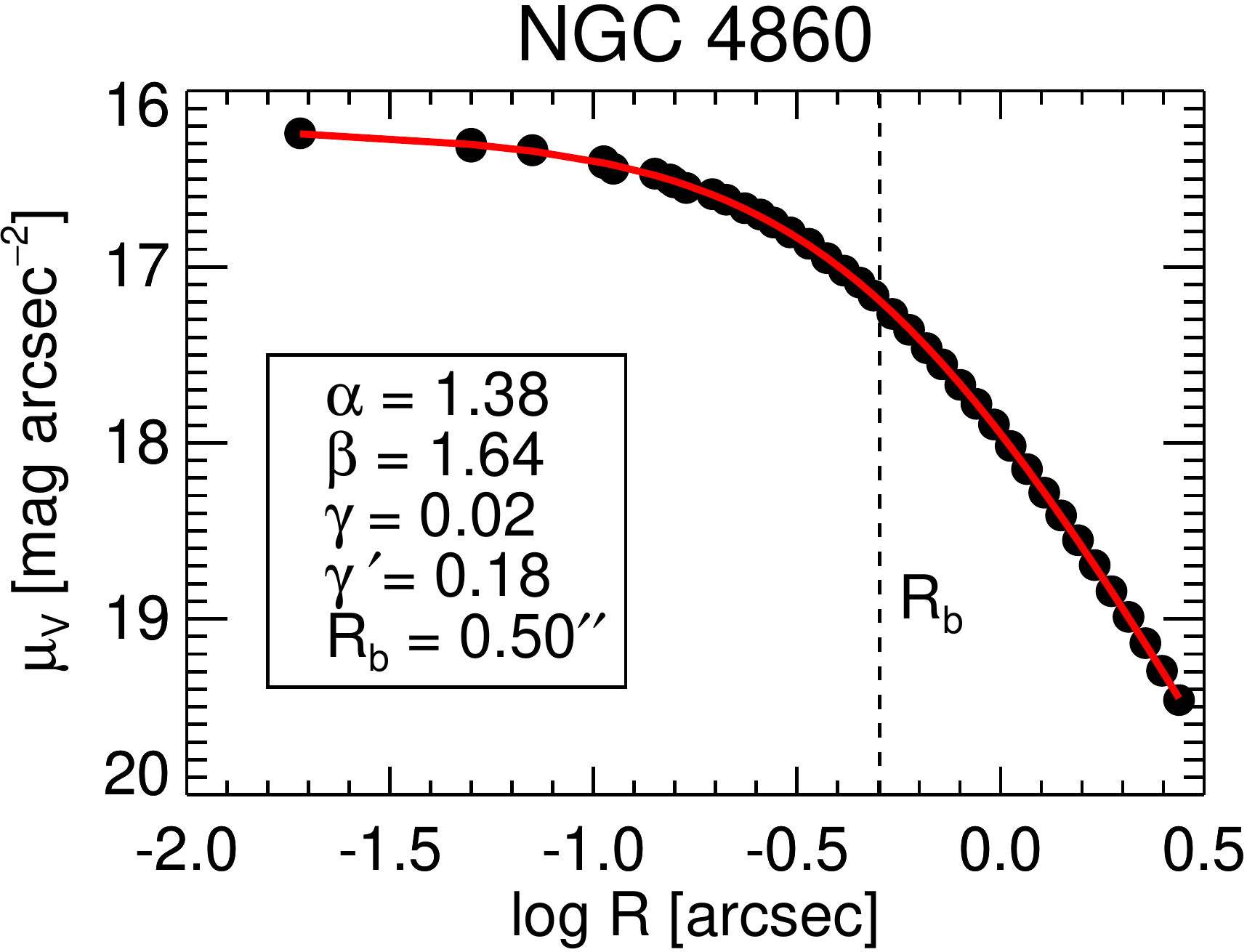}
\caption{Nuker-law fit to the surface brightness profile of the Coma slow rotator NGC~4860. The surface brightness profile of the galaxy (filled circles) is plotted against the radius. The best fitting relation is overlaid with the red solid line and the corresponding parameters are written in the caption. The logarithmic slope  $\gamma'=0.18<0.3$ at $R=0\farcs1$ identifies NGC~4860 as a core galaxy.}
\label{fig:profile_ngc4860}
\end{figure}

\begin{figure*}
\plotone{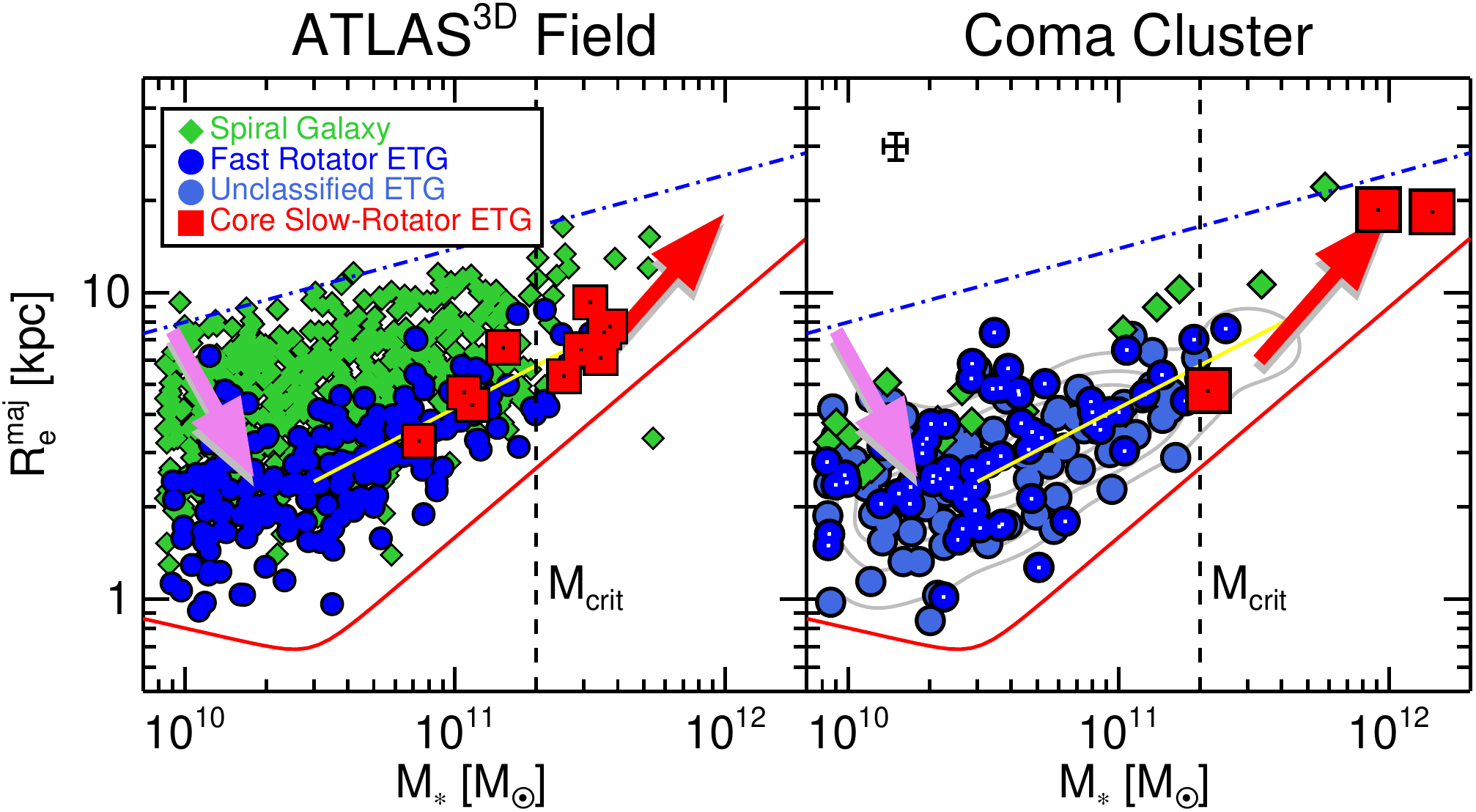}
\caption{Mass-size distributions in extreme environments. The left panel shows the field subsample of the \atl\ sample of nearby galaxies. The right panel shows an identically-selected sample of Coma cluster members. A representative error bar is indicated. The gray contours in the right panel show for reference the kernel density estimate for the ETGs distribution in the left panel, while the yellow line is the corresponding mean relation (from fig.~2 of \citealt{Cappellari2013p20}).  For references the blue dash-dotted line and the thick red line indicate the upper limit of the spiral galaxies and the lower limit for ETGs from \citet{Cappellari2013p20}. The critical mass $M_{\rm crit}$ is marked by the black dashed line. The magenta arrow qualitatively indicates the evolutionary track due to bulge growth and quenching, with little mass increase. The red arrow shows the models prediction track for major dry merging with a factor 3 mass increase.
}
\label{fig:mass_size}
\end{figure*}

\subsection{Galaxy sizes and masses}

All galaxy sizes are homogeneously taken from the XSC. Effective radii \remaj\ are defined as the major axis of the isophote enclosing half of the {\em total} galaxy light in $J$-band (XSC keyword j\_r\_eff) which has better $S/N$. The use of \remaj\ is needed to remove the strong inclination dependence of the circularized \re, for disk galaxies \citep{Cappellari2013p15}. The XSC effective radii are among the most reproducible relative size measures \citep{Cappellari2013p15}. However the absolute normalization of \remaj\ depends on the quality of the data. We define $\remaj=1.61\times$j\_r\_eff to match the \remaj\ of \citet{Cappellari2013p15} for the galaxies in common, making our results directly comparable.

Sizes measured via growth curves saturate near the FWHM of the 2MASS PSF, which is $\sim1.5$ kpc at Coma. We corrected j\_r\_eff using the following formula which we found via simulations is 5\% accurate for  an $R^{1/4}$ profile, a Gaussian PSF, and $R_{\rm e}^{\rm true}>\sigma_{\rm PSF}/2$, with weak dependence on the Sersic index and flattening
\begin{equation}
     (R_{\rm e}^{\rm true})^2 = 
    (R_{\rm e}^{\rm obs})^2 - (1.45\times\sigma_{\rm PSF})^2.
\end{equation}
This is approximate but is not critical for our conclusions as it only affects the few smallest Coma galaxies.

Galaxy masses derived from stellar population necessarily ignores systematic variations of about a factor of two in the stellar Initial Mass Function \citep[e.g.][]{Cappellari2012,Conroy2012}. For this reason our masses are based on $K_s$-band luminosities:
\begin{equation}
\log_{10} M_\ast\approx10.58-0.44\times(M_{K_s}+23).
\label{eq:mk_to_mjam}
\end{equation}
This was fitted (\reffig{fig:from_kmag_to_mass}) using the LTS\_LINEFIT routine of \citet{Cappellari2013p15} to the \atl\ dynamical masses $M_{\rm JAM}$ from \citet{Cappellari2013p15}. We then assumed $M_\ast=M_{\rm JAM}$ as discussed in sec.~4.3 of \citet{Cappellari2013p15}. Our relation is fitted to ETGs and may become inaccurate for spirals. However \citet{Williams2009} find the $(M/L)_{K_s}$ of spirals and ETGs does not differ by more than $\sim50\%$. None of our conclusions depend on this approximation.

\section{Recognizing merging history}

We want to robustly recognize genuine spheroidal-like early-type galaxies that are the likely result of dry mergers from inclined disk-like systems. Integral-field stellar kinematics was shown to provide an excellent discrimination of these two classes of galaxies, nearly independently of inclination. The two classes were called slow and fast-rotator ETGs respectively \citep{Emsellem2007,Cappellari2007}. 

Another signature which was shown to indicate dry merger remnants is the presence of a core or light deficit in the inner surface-brightness profile \citep[e.g.][]{Kormendy2009}. Although the kinematic approach is more robust, in many cases the core and slow-rotator classifications agree as expected \citep{Lauer2012}. However there are important cases where either core galaxies or slow-rotator ETGs appear clearly disk-like and inconsistent with being dry-merger remnants \citep{Krajnovic2013p23}. We found that selecting only slow rotators {\em with core} eliminates from the class the `misclassified'  flat counter-rotating disks or unsettled mergers \citep{Krajnovic2011,Emsellem2011} and appears to unambiguously indicates dry-merger remnants only. For this reason here we use core slow-rotator galaxies to define dry mergers remnants.

For the \atl\ sample we take the fast/slow rotator separation from \citet{Emsellem2011} and the cusp/core one from \citet{Krajnovic2013p23}. For the Coma sample we used the fast/slow rotator classes recently derived from integral-field spectroscopy (IFS) by \citet{Houghton2013}. The three slow rotators ($C=3$ in their table~1), NGC~4874, NGC~4884 ($=$NGC~4889) and NGC~4860, all have a core. For the first two objects the core classification is given in \citet{lauer07prof}, and we classified the third one from HST photometry (\reffig{fig:profile_ngc4860}). Unlike for the \atl\ sample, the Coma IFS observations are not complete but cover 27 ETGs representative of the population within the innermost 15 arcmin from the cluster center. We classified 39 additional objects as fast rotator from their apparent flattening: all galaxies flatter than $\varepsilon>0.4$ (from XSC keyword sup\_ba) must be fast rotators as shown by \citet{Emsellem2011}. In the next section we discuss why this subsample allows us to reach general conclusions for the entire cluster.

\begin{figure*}
\plotone{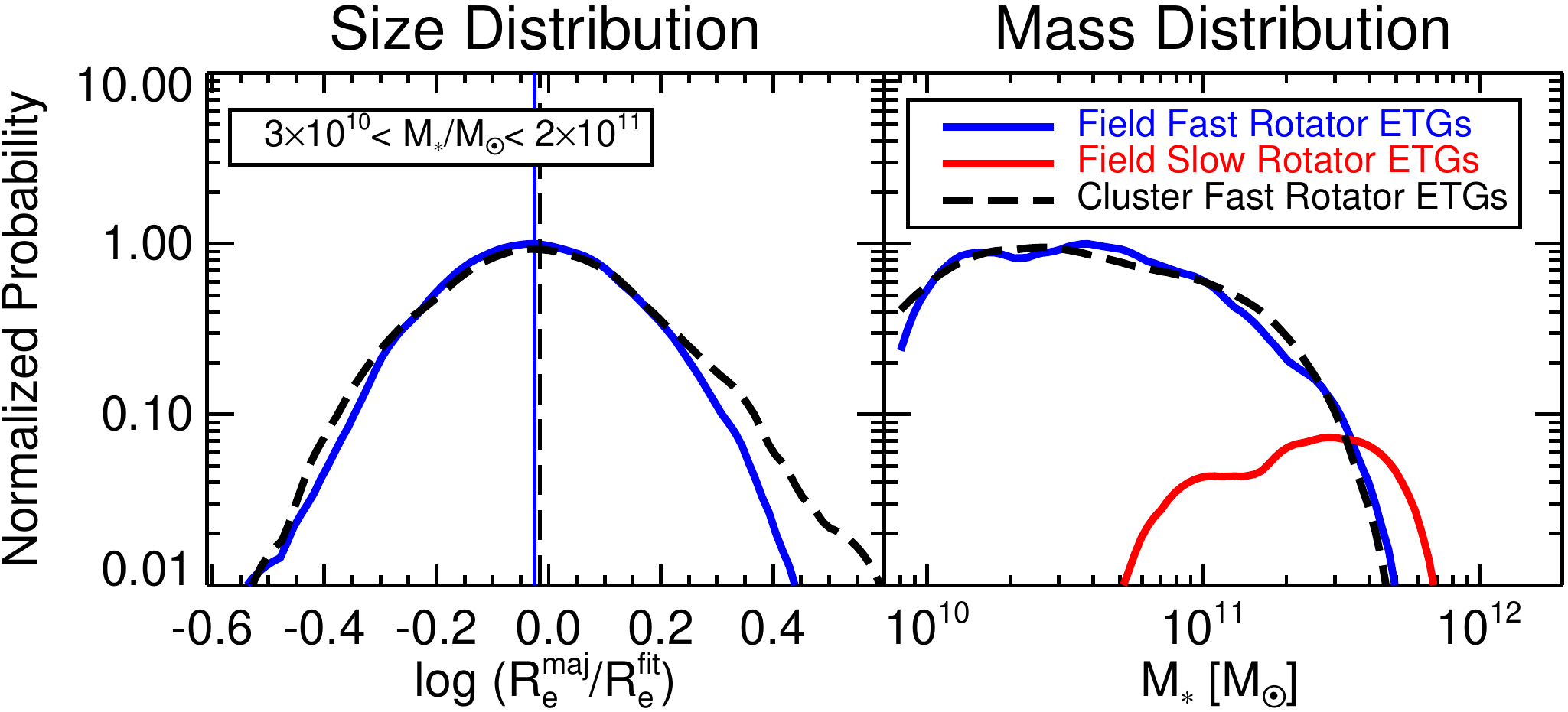}
\caption{Size and mass distributions of fast rotator ETGs in the field and cluster samples. In the left panel a normalized kernel density estimate of the probability distribution, is plotted as a function of the logarithm of the ratio between the observed $R_{\rm e}^{\rm maj}$ and the value predicted by the best-fitting relation $(R_{\rm e}^{\rm fit}/{\rm kpc})=4.2\times[M_\ast/(10^{11}\msun)]^{0.46}$ (yellow line in \reffig{fig:mass_size}). This is done to remove the significant size variation within the selected mass interval $3\times10^{10}<M_\ast/\msun<2\times10^{11}$. The vertical lines indicate the biweight mean of the distributions which differ by just 4\%. The tail at larger sizes in the cluster sample is likely due to red spirals, which are more common in clusters and are easily confused with ETGs. This asymmetry may explain some of the reported size increase of passive galaxies in denser environments. The right panel shows the very similar distributions of galaxy masses. The core slow-rotators in the field are shown in red.}
\label{fig:mass_distribution}
\end{figure*}

\section{Mass-size relation in extreme environments}

The mass-size distribution for the field subset of the \atl\ parent sample is presented in the left panel of \reffig{fig:mass_size} (for the entire sample see fig.~9 of \citealt{Cappellari2013p20}). It shows two nearly parallel sequences of spiral galaxies and fast-rotator ETGs, with the latter having smaller size at given mass.
Core slow-rotators do not follow the trends of spirals and fast rotator ETGs. They lie along the mass-size relation defined by fast rotators but  are only present above $M_\ast\ga10^{11}$ \msun\ \citep{Krajnovic2013p23} and start dominating the ETGs population above $M_{\rm crit}\approx2\times10^{11}$ \msun, where a number of galaxy properties abruptly change \citep{Cappellari2013p20}.

\begin{figure*}
\centering
\includegraphics[height=8cm]{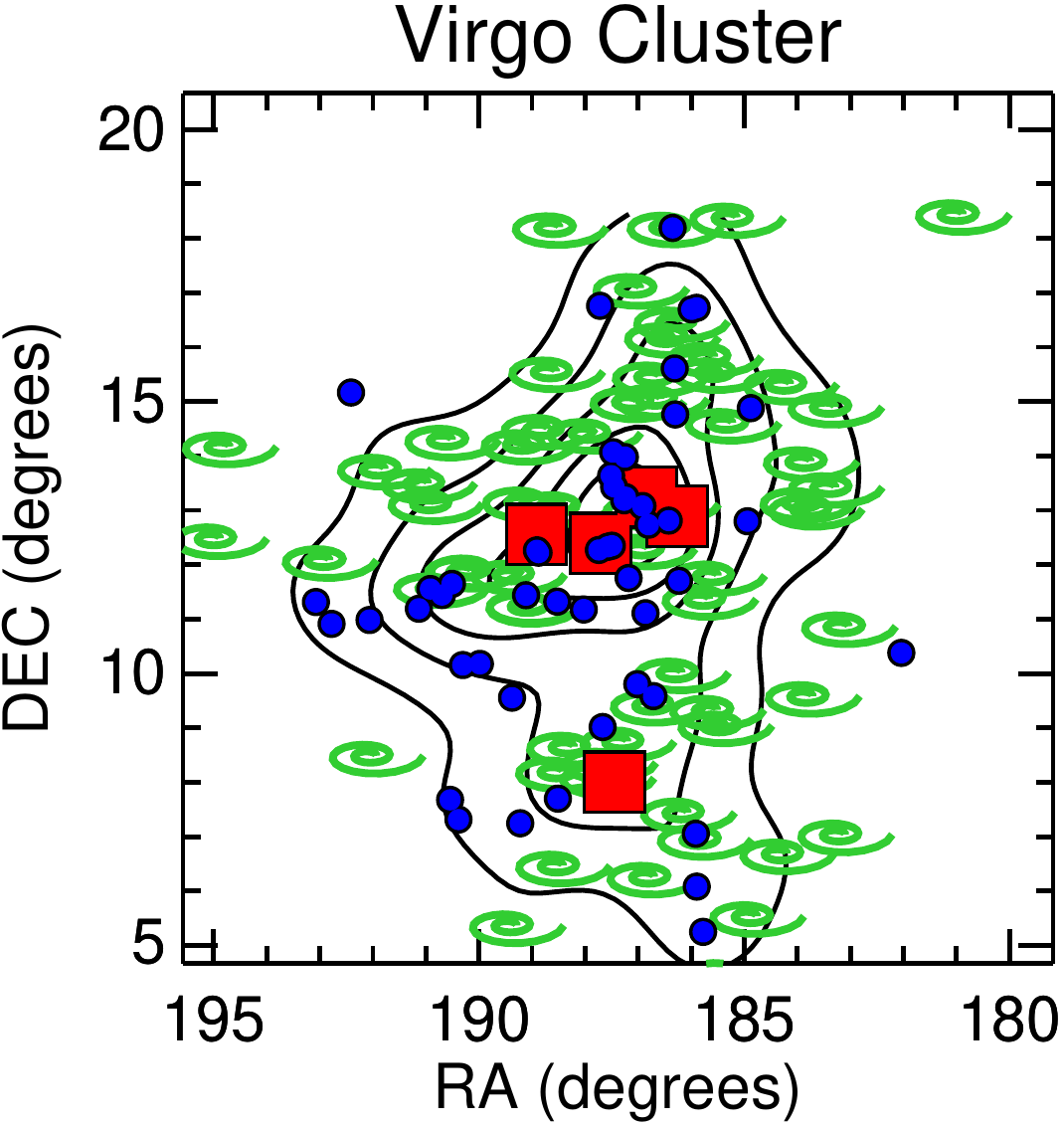}
\includegraphics[height=8cm]{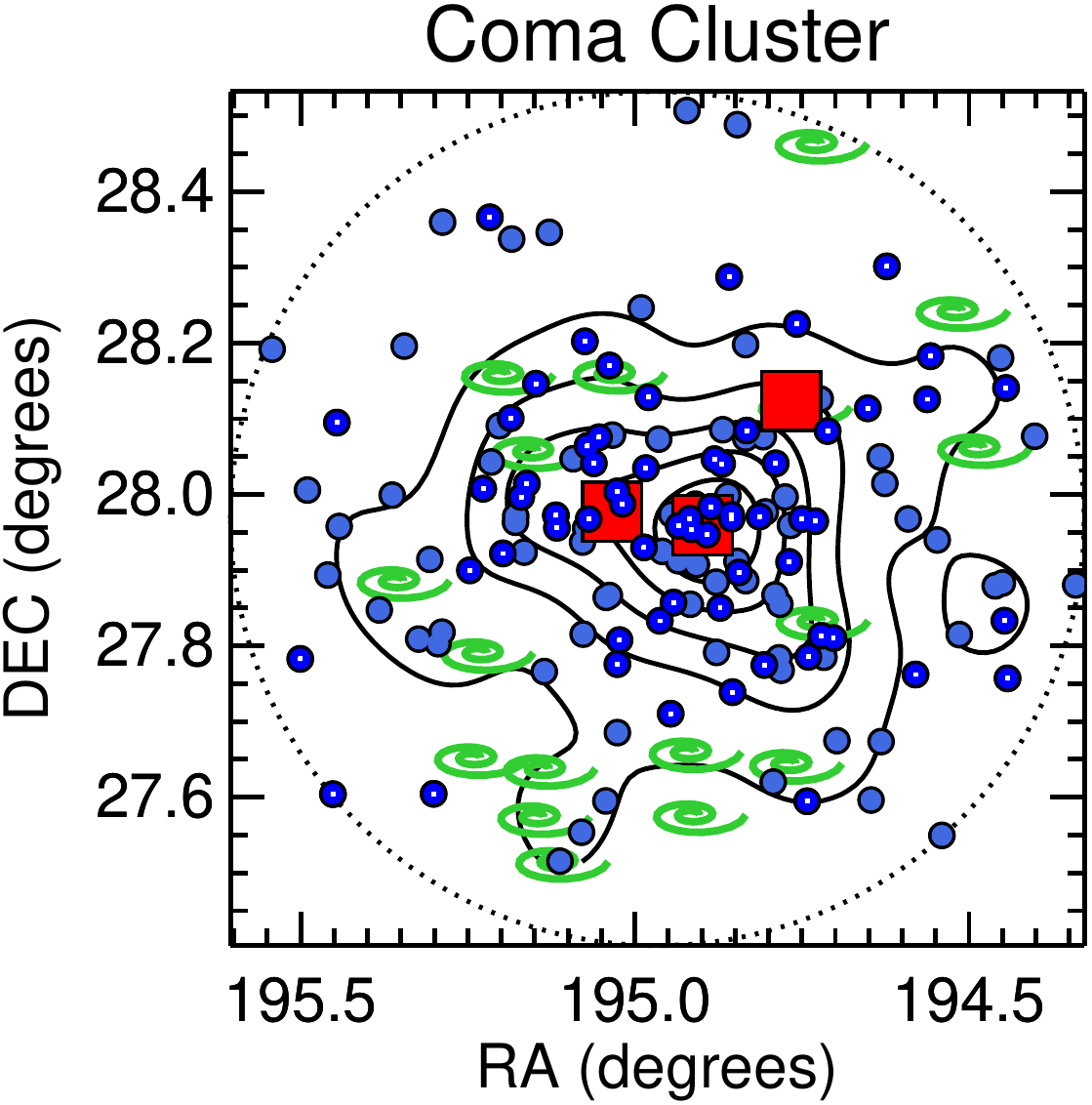}
\caption{Distribution of galaxies in the Virgo and Coma clusters. Symbols are the same as in \reffig{fig:mass_size}, except that here the spirals are shown as green spirals. The large dotted circle in the right panel is our selection limit. A kernel density estimate for the galaxy distribution is overlaid with linearly spaced contours. The spirals and fast-rotator ETGs distribution is quite different in the two panels, but in both cases the few core slow-rotators are concentrated near the density peaks.}
\label{fig:coma_map}
\end{figure*}

The mass-size relation for the galaxies in the Coma cluster is shown in the right panel of \reffig{fig:mass_size}. We also quantify the cluster/field mass and size distributions in \reffig{fig:mass_distribution}. The following results are obvious: (i) spiral galaxies in Coma are replaced by fast rotator ETGs which follow the {\em same} mass-size relation and have the {\em same} mass distribution as the field sample; (ii) slow rotator ETGs in Coma lie above $M_{\rm crit}$ and the two largest galaxies appear segregated in mass from the fast rotators, with their size increasing proportionally to their mass. The two massive slow rotators stand out for sitting near the center of the cluster, while the third one lies along a slight overdensity (\reffig{fig:coma_map} right).

Although the Coma IFS observations do not sample the entire cluster, they do sample most of its densest part (\reffig{fig:coma_map} right). The fact that no core slow rotator was found below $M_{\rm crit}$ in the densest parts, makes it unlikely that others may be found in the rest of the cluster. This is because previous IFS studies in the Virgo (\reffig{fig:coma_map} left; \citealt{Cappellari2011b}), Abell 1689 \citep{DEugenio2013} and Coma cluster \citep{Houghton2013} have shown core slow rotators to be strongly concentrated towards the densest part of the clusters. Even making the incorrect and extreme assumption that the distribution of slow rotators is independent of environment, using the hypergeometric distribution we can say that if there were more than just three slow rotators among the ETGs without IFS, we would have a $>50\%$ chance of observing at least one. We find no slow rotator below $M_{\rm crit}$.

\section{Discussion}

\subsection{Two formation processes}

Our work relies on our ability to distinguish, within the ETGs class, the relics of dry mergers, the core slow-rotators, from inclined passive disks with a range of bulge fractions, the fast rotators. The comparison between the mass-size relation we observe in the field and in the core of the dense Coma cluster, reveals two distinct processes transforming galaxies in clusters: (i) Spirals transform into fast rotator ETGs while decreasing in \remaj\ with little mass variation, while (ii) slow rotators increase in \remaj\ roughly proportionally to their $M_\ast$, segregating in mass from the fast rotators.

As discussed in \citet{Cappellari2013p20}, the decrease of \remaj\ from the spirals to the fast rotator ETGs traces the bulge growth, concentrating mass at smaller radii. The fact that fast rotator ETGs have smaller \remaj\ than spirals shows that the environmental quenching of star formation is associated to the bulge growth (and likely some disk fading). The fact that the mass distribution of fast rotators is the same in both the field and in Coma shows that mass increased insignificantly during the environmental transformation \citep[see also][]{Carollo2013}. 

Various processes can transform spirals into passive ETGs  \citep[see][for a review]{Boselli2006}. Both high-speed encounters within the cluster or secular evolution can drive gas towards the center, growing a bulge and producing a starburst which is subsequently quenched by some feedback (e.g. AGN or supernovae). These will combine with ram-pressure stripping of the cold gas or other forms of more gradual gas starvation. 

An opposite alternative to explain the differences in the field and cluster mass-size relations would be the lack of disk growth around pre-existing spheroids in the cluster environment. The homogeneity in the dynamical structure of bulges and disks of fast rotators \citep{Cappellari2013p15}, the similarity in the maximum mass of fast-rotator ETGs and spirals, and the evolution of the surface brightness profiles as a function of redshift \citep{vanDokkum2013} seem to rule out this scenario.

The situation is completely different for the slow-rotator ETGs. Compared to the field sample, the high-density environment appears to evolve these objects along lines of $\remaj\propto M_\ast$ in the mass-size plane. The two most massive slow rotators in Coma appear segregated in mass from the fast rotator population, with a $\sim5\times$ gap in $M_\ast$, which is not present in the field sample. The median mass increase for the slow rotators in Coma with respect to our field sample is about a factor three.

The size increase in approximate proportion to the mass, combined with the mass segregation of slow rotators from fast rotators in Coma, is direct evidence for major growth of slow rotators by dry mergers in the cluster environment.  The observed size increase with mass is the one predicted for major (not minor) dry mergers. This is consistent with the mass versus velocity dispersion relation of nearby ETGs \citep{Cappellari2013p20,Kormendy2013}. However very massive galaxies generally have  extended stellar halos which may be missed by 2MASS. Deeper observations would likely indicate larger \re\ \citep[e.g.][]{Kormendy2009}, making the size increase more consistent with the minor merging formation hypothesis \citep[e.g.][]{Naab2009}.

The hierarchical paradigm for galaxy formation implies that galaxies experience a variety of environments during their evolution. Core slow-rotators do not need to form in clusters, as illustrated by our field sample. They likely form in efficient starburst in the high-redshift Universe and, due to their large masses, they sink to the center of groups via dynamical friction. When groups merge to form massive clusters, slow rotators again sink towards the center where they merge to form more massive slow rotators. The same cannot happen to fast rotators, which have too small masses to efficiently sink to the center and too fast velocities to merge. They are quenched by the cluster but essentially don't change their mass. This picture is broadly consistent with theoretical studies \citep[e.g.][]{DeLucia2012}. 

A progenitor of the Coma cluster may have looked like the less massive Virgo cluster which contains four core slow-rotators closely packed within its central core (subcluster A), surrounded by a swarm of fast rotators and a much larger fraction of spirals than in Coma (\reffig{fig:coma_map}). Another core slow rotator lies at the center of Virgo subcluster B \citep[see][]{Cappellari2011b}. Consistently with this scenario is the fact that the Virgo core slow-rotators are more numerous than in Coma and not yet segregated in mass from the fast rotators.

\subsection{Explaining mass, environment and redshift trends}

The existence of two distinct evolutionary paths for fast and slow rotator ETGs naturally explains a number of previously reported empirical trends, which seems different manifestations of this phenomenon.

Our result explains why galaxy quenching appears to depend in a separable way on mass and environment \citep{Peng2010,Smith2012coma}. \reffig{fig:mass_size} shows that the ``environment quenching'' consist of the transformation of spirals into fast rotator ETGs with similar masses, while ``mass quenching'' is a combination of the two facts that: (i) when spirals transform into fast rotators, quenching is driven by bulge mass fraction, irrespective of environment  \citep{Cappellari2013p20}, and (ii) when slow rotator ETGs grow in mass they quench above $M_{\rm crit}\approx2\times10^{11}$ \msun, irrespective of environment. 

Two distinct processes, ``bulge-driven'' and a ``halo-driven'' quenching, were recently proposed to explain the distribution of galaxy properties, stellar population and gas content on the mass-size relation of the \atl\ sample \citep{Cappellari2013p20} and the link between star formation and central density in galaxies \citep{Cheung2012}. Our study confirms and clarify the nature of the two processes involved.

The different formation paths of fast and slow rotator ETGs described previously, appears related with the different ``outside-in'' versus ``inside-out'' redshift evolution of sizes and profiles of galaxies with masses below/above $M_{\rm crit}$ \citep[e.g.][]{vanDokkum2010profiles,vanDokkum2013}. Our results suggests one should identify the first class with the progenitors of fast rotators and the second one with the progenitors of the slow rotator ETGs.

The lack of environmental size variation we observe for fast rotators and the strong size increase of slow rotators explains some contrasting results on the environmental size dependence, as this depends on the fraction of fast and slow rotators in the samples.

The bulge growth versus dry merging distinction also explains the fact that, while the ratio of fast rotators to spirals increases strongly with number density, there is no clear trend in the ratio of slow rotators to fast rotators, except in the center of clusters \citep{Cappellari2011b}. This result was shown to hold up to the densest environments \citep{Scott2012,DEugenio2013,Houghton2013}.

\acknowledgements

I thank the referee for a very useful report. I acknowledge support from a Royal Society University Research Fellowship.

\end{document}